\documentclass[aps,pra,
onecolumn,twoside]{revtex4-2}
\usepackage[smalltableaux]{ytableau}
\usepackage{float,lipsum,enumitem,graphicx,csquotes,tikz,makecell,setspace,pifont,hyperref,moresize}
\usepackage{bbm,bm,mathrsfs,amsmath,amsthm,amssymb,mathtools,dsfont,mathdots,stmaryrd}

\newcommand{\nc}{\newcommand}				\newcommand{\rnc}{\renewcommand}
\nc{\x}{\textnormal} \nc{\n}{\operatorname} \nc{\s}{\mathsf} \nc{\bb}{\mathbb}
\nc{\alp}{\alpha}  \nc{\bt}{\beta}			\nc{\gm}{\gamma}  \nc{\Gm}{\Gamma} \nc{\dt}{\delta}
\nc{\Dt}{\Delta}   \nc{\kp}{\kappa}			\nc{\sg}{\sigma}  \nc{\Sg}{\Sigma} \nc{\tht}{\theta}
\nc{\Tht}{\Theta}  \nc{\ld}{\lambda}		\nc{\Ld}{\Lambda} \nc{\om}{\omega} \nc{\Om}{\Omega}
\nc{\phv}{\varphi} \nc{\epsl}{\varepsilon}	\nc{\thv}{\vartheta}
\nc{\Cal}[1]{\mathcal{#1}} \nc{\fr}[1]{\mathfrak{#1}}
\nc{\Ac}{\Cal{A}} \nc{\Bc}{\Cal{B}} \nc{\Cc}{\Cal{C}} \nc{\Dc}{\Cal{D}} \nc{\Ec}{\Cal{E}}
\nc{\Fc}{\Cal{F}} \nc{\Gc}{\Cal{G}} \nc{\Hc}{\Cal{H}} \nc{\Ic}{\Cal{I}} \nc{\Jc}{\Cal{J}}
\nc{\Kc}{\Cal{K}} \nc{\Lc}{\Cal{L}} \nc{\Mc}{\Cal{M}} \nc{\Nc}{\Cal{N}} \nc{\Oc}{\Cal{O}}
\nc{\Pc}{\Cal{P}} \nc{\Qc}{\Cal{Q}} \nc{\Rc}{\Cal{R}} \nc{\Sc}{\Cal{S}} \nc{\Tc}{\Cal{T}}
\nc{\Uc}{\Cal{U}} \nc{\Vc}{\Cal{V}} \nc{\Wc}{\Cal{W}} \nc{\Xc}{\Cal{X}} \nc{\Yc}{\Cal{Y}}
\nc{\Zc}{\Cal{Z}}
\nc{\im}{\n{im}} \nc{\Tr}{\n{Tr}} \nc{\tr}{\n{tr}} \nc{\rank}{\n{rank}} \nc{\rk}{\n{rk}}
\nc{\Bb}{\bb{B}} \nc{\Cb}{\bb{C}} \nc{\Fb}{\bb{F}} \nc{\Hb}{\bb{H}} \nc{\Nb}{\bb{N}}
\nc{\Rb}{\bb{R}} \nc{\Sb}{\bb{S}} \nc{\Zb}{\bb{Z}}
\rnc{\P}{\bb{P}} \nc{\E}{\n{\bb{E}}} \nc{\Eb}{\mathop{{}\bb{E}}}
\nc{\su}{\fr{su}} \nc{\gf}{\fr{g}}  \nc{\hf}{\fr{h}} 
\nc{\RR}{\mathbf{R}} \nc{\id}{\x{id}} \nc{\SU}{\x{SU}} \nc{\Ux}{\x{U}}
\nc{\poly}{\x{poly}} \nc{\Hom}{\x{Hom}}

\nc{\ot}{\otimes} \nc{\Ot}{\bigotimes} \nc{\Oplus}{\bigoplus}
 \nc{\PT}[1]{^{\s{T}_{\!{#1}}}}
\nc{\dg}{^{\dagger}} \nc{\f}[2]{\frac{#1}{#2}}
\nc{\bra}[1]{\langle{#1}|}
\nc{\ket}[1]{|{#1}\rangle}
\nc{\brak}[1]{\langle{#1}\rangle}			\nc{\ketb}[2]{|{#1}\rangle\!\langle{#2}|}
\nc{\xto}[1]{\xrightarrow{#1}} \nc{\dd}{\,\x{d}} \nc{\ee}{\x{e}} \nc{\ii}{\x{i}} 
\nc{\1}{\mathds{1}}  \nc{\2}{\{0,1\}}		\nc{\lims}{\varlimsup}  \nc{\limi}{\varliminf}
\nc{\ty}[1]{{\x{\tiny $#1$}}}				\nc{\too}{\!\!\to\!\!}

\nc{\be}[2]{\begin{#1}#2\end{#1}}
					\nc{\Thm}[1]{\be{thm}{#1}}
			\nc{\Prop}[1]{\be{prop}{#1}}
			\nc{\Coro}[1]{\be{coro}{#1}}
				\nc{\Lem}[1]{\be{lem}{#1}}
\nc{\Pf}[1]{\be{proof}{#1}}
\nc{\Eq}[1]{\be{equation}{#1}} \nc{\Eqn}[1]{\be{equation*}{#1}}
\nc{\Al}[1]{\be{align}{#1}} \nc{\Aln}[1]{\be{align*}{#1}} \nc{\Als}[1]{\be{align}{\be{split}{#1}}}
\nc{\Item}[1]{\be{itemize}{#1}} \nc{\Enum}[1]{\be{enumerate}{#1}}
\nc{\gray}[1]{{\color{gray}#1}} \nc{\red}[1]{{\color{red}#1}} \nc{\blue}[1]{{\color{blue}#1}}
\nc{\mm}[2]{\ps{\begin{array}{c}{#1}\\{#2}\end{array}}}
\nc{\mmnn}[4]{\ps{\begin{array}{cc}{#1}&{#2}\\{#3}&{#4}\end{array}}}

\begin{document}
\title{Ancilla-free certification of unitary quantum processes}
\author{Wei Xie}
\email{xxieww@ustc.edu.cn}
\affiliation{School of Computer Science and Technology, University of Science and Technology of China}
\affiliation{The CAS Key Laboratory of Wireless-Optical Communications, University of Science and Technology of China}

\begin{abstract}
We study eﬀicient quantum certification algorithms for unitary quantum process using no ancilla. Previous study showed that one can distinguish whether an unknown unitary $U$ is equal to or $\varepsilon$-far from a known or unknown unitary $V$ in fixed dimension with $O(\varepsilon^{-2})$ uses of the unitary, in which the Choi state is used and thus a high dimensional ancilla system is always needed. We give an algorithm that distinguishes the two cases with $O(\varepsilon^{-1})$ uses of the unitary, using fewer or no ancilla, outperforming previous relevant results.
\end{abstract}

\date{\today}
\maketitle

\section{Introduction}



A fundamental task in building reliable quantum information processing devices is to obtain parameters of an unknown quantum device.
This is called {\em tomography} if all parameters are required to be known.
In many scenarios, however, we are concerned only with whether the unknown state or operation satisfies specific property.
For example, to assess the quality of a quantum chip after production, one needs to check whether the circuit is close to a given unitary transformation, and it is unnecessary to get all parameters about this chip.
This is called {\em certification}, which usually saves state copies, query numbers, or storage space compared with the quantum tomography.
See \cite{montanaro2016survey} for a survey on quantum certification.


The task of quantum property testing of unitary quantum processes is specified by two disjoint sets $\Pc,\Qc$ of processes.
Given the classical discription of the two sets and access to an unknown unitary $U$, a tester (i.e., a quantum algorithm) $\Tc$ should either accept (i.e., reports $U\in\Pc$) or reject (i.e., reports $U\in\Qc$), promised one is the case.
The tester $\Tc$ is {\em eligible} if the following conditions are satisfied:

(1) (Completeness) If $U\in \Pc$, $\Tc$ accepts with probability at least 2/3;

(2) (Soundness) If $V\in\Qc$, $\Tc$ accepts with probability at most 1/3.

If the tester accepts with certainty in the case $U\in\Pc$, we say the tester has {\em perfect completeness}.
The goal is to design an eligible tester which uses as few queries to the process as possible, and the minimum number of queries is called the {\em query complexity} of this task.
The numbers $2/3,1/3$ in the completeness and soundness conditions can be replaced by any constants $c,s$ satisfying $1>c>1/2>s>0$, respectively, which only changes the copy complexity by constant factors due to the confidence amplification by repeating the test.
Similar to quantum state certification, in the task of quantum unitary certification, $\Pc=\{\ee^{\ii\theta}V:\theta\in\Rb\}$ for some unitary $V$ and $\Qc=\{W\x{ is a unitary}: \x{dist}(W,V)>\epsl\}$ for some distance measure dist and some small positive number $\epsl$.
For the task quantum unitary certification for two unknown unitaries, $\Pc=\{(U,V):\x{dist}(U,V)=0\}$ and $\Qc=\{(U,V):\x{dist}(U,V)>\epsl\}$.



In order to test whether a pair of unknown states $\rho$ and $\sg$ on $\Hc$ are equal, the swap test \cite{buhrman2001quantum} is usually used.
By generalizing the swap test and applying group representation theory, it is shown in \cite{buadescu2019quantum} that $O(d/\epsl^2)$ copies of quantum states suffices to distinguish whether an unknown $\rho$ is equal to some known $\sg$ or $\epsl$-far from $\sg$ in trace distance.
This method is efficient since the tomography needs $\Om(d^2)$ copies of quantum states.


The certification of unitaries is quite different from that of quantum states in that the quantum unitary certification requires a double optimization, one is the input state for the quantum unitary and the other is the choice of measurement after the action of quantum unitary.
The works in \cite{da2011practical,steffen2012experimental,reich2013optimal} used methods based on Monte Carlo sampling to estimate the fidelity of an unknown gate to a fixed one, and also studied the optimality of estimation strategy.
Given an unknown unitary $U$ and a known or unknown unitary $V$, by using the Choi correspondence between quantum unitaries and states, there exists a tester that distinguishes whether their distance is zero or larger than $\epsl$ with $O(\epsl^{-2})$ uses of unitaries.
By using the Schur-Weyl decomposition, we show that for fixed dimension of the unitary, only $O(\epsl^{-1})$ uses of the unitaries suffice to achieve the same goal.
Another advantage of our algorithm is that we do not need to introduce extra ancilla system as in the method using Choi state.

\section{Testing equality of unitaries}\label{sec4.3}


Following \cite{montanaro2016survey}, define the distance between two unitaries $U,V\in\Ux(d)$ as
\Eq{\label{defjfogbben}
\x{dist}(U,V)=\sqrt{1-\Big|\f{1}{d}\brak{U,V}\Big|^2} \,,
}
where $\brak{U,V}:=\tr(U\dg V)$.
The distance of two unitaries is no larger than 1.

Notice that the normalized inner product of unitaries is equal to the inner product of their Choi states, i.e.,
\Eq{
\f{1}{d}\brak{ U,V }=\bra{\phi^+} (U\dg\ot\1)(V\ot\1) \ket{\phi^+} \,.
}
Here and in the following we use $\ket{\phi^+}$ to denote the (normalized) maximally entangled state.
Consider the Schur-Weyl decomposition
\Eqn{
	(\Cb^d)^{\ot n} = \Oplus_{\ld\vdash(n,d)} \Hc_\ld\ot\Kc_\ld \,,
}
where $\Hc_\ld$ are irreducible representation spaces of $\Ux(d)$ and $\Kc_\ld$ are multiplicity spaces.
Here $\ld\vdash(n,d)$, also written $\ld\in\x{Par}(n,d)$, means that $\ld$ is a partition of $n$ with length at most $d$.

The dimension of $\Kc_\ld$, for partition $\ld$ of $n$, is given by
\Eq{\label{fjoiwhoggefho}
\dim \Kc_\ld = \f{n!}{\widetilde{\ld}_1!\cdots\widetilde{\ld}_d!}\prod_{1\le i<j\le d}\big(\widetilde{\ld}_i-\widetilde{\ld}_j\big) \,,
}
where $\widetilde\ld_i:=\ld_i+d-i$.
It can be bounded by \cite{hayashi2002exponents}
\Eq{
\binom{n}{\ld}(n+d)^{-d(d-1)/2} \le \dim V_\ld^S \le \binom{n}{\ld}
}
and furthermore
\Eq{\label{vmognreogbjgt2}
	\exp(nH(\bar \ld)) (n+d)^{-d(d+1)/2} \le\dim V_\ld^S \le \exp(nH(\bar \ld)) \,,
}
where $\bar\ld:=\ld/n$.

The character $\chi_\ld^L$ for the irrep $\Hc_\ld$ is
\Eq{\label{owdhbeobhro}
\chi_\ld^L(\x{diag}(x_1,\dots,x_d))=\f{\det(x_i^{\ld_j+d-j})_{i,j=1}^d}{\det (x_i^{d-j})_{i,j=1}^d} \,.
}
The dimension of $\Hc_\ld$ is given by
\Eq{\label{vmognreogbjgt}
\dim \Hc_\ld = \prod_{1\le i<j \le d}\f{\ld_i-\ld_j+j-i}{j-i} \,,
}
which is bounded above by $(n+1)^{d(d-1)/2}$ \cite{christandl2006spectra}.

The entanglement between the representation space and multiplicity space was used to estimate the group transformation \cite{chiribella2005optimal}.
It turns out that it can be also used in the certification of unitaries as shown in the following.

\Thm{\label{thm:vnewrhod}
	Given access to an unknown single-qubit unitary $U$, there exists a tester with perfect completeness that distinguishes whether $U$ is equal to a fixed and known $V$ up to a phase or $\x{dist}(U,V)\ge\epsl$ with $O(\epsl^{-1})$ uses of $U$, without using any ancilla system.
}
\begin{proof}
By Schur-Weyl duality, $(\Cb^2)^{\ot n}=\Oplus_j \Hc_j\ot\Kc_j$ where $\Hc_j$ is the irrep of $\Ux(2)$ of dimension $2j+1$ and $\Kc_j$ is the corresponding irrep of $S_n$.
In this decomposition we write $U^{\ot n}=\Oplus_j U_j\ot\1_j$ for any $U\in\Ux(2)$ since $U^{\ot n}$ acts as identity operator on each $\Kc_j$.

Notice that $\Hc_{\f{n}{2}-1}$ and $\Kc_{\f{n}{2}-1}$ have the same dimension $n-1$.
Let $\ket{\phi^+}$ be the maximally entangled state in $\Hc_{\f{n}{2}-1}\ot\Kc_{\f{n}{2}-1}$, i.e., $\ket{\phi^+}=\f{1}{\sqrt{n-1}}\sum_{i=1}^{n-1}\ket{\alp_i}\ket{\bt_i}$ where $\{\ket{\alp_i}\}$ and $\{\ket{\bt_i}\}$ are orthonormal bases of $\Hc_{\f{n}{2}-1}$ and $\Kc_{\f{n}{2}-1}$ respectively.

Apply the unitary $U^{\ot n}$ to $\ket{\phi^+}$, and then perform the POVM $\{ \phi^+_V,\1-\phi^+_V \}$ where $\ket{\phi^+_V}:=(V_{\f{n}{2}-1}\ot\1_{\f{n}{2}-1})\ket{\phi^+}$.
The tester reports that $U$ equals $V$ up to a phase if the first outcome occurs, and reports $\x{dist}(U,V)\ge\epsl$ otherwise.
Obviously the tester has perfect completeness irrespective of the choice of $n$.
Next step is to derive the requirement for $n$ to ensure the soundness condition $\brak{ U^{\ot n}\phi^+U^{\dagger,\ot n},\phi^+_V }\le1/3$ when $\x{dist}(U,V)\ge\epsl$.

Denote the eigenvalues of $U\dg V$ by $\ee^{\ii\alp}$ and $\ee^{\ii\bt}$ for $\alp,\bt\in[0,2\pi)$.
When $\x{dist}(U,V)\ge\epsl$, then by the definition~\eqref{defjfogbben}, $|\brak{U,V}|^2\le4(1-\epsl^2)$, i.e., $|\ee^{\ii\alp}+\ee^{\ii\bt}|^2\le4(1-\epsl^2)$.
It follows that $|\sin\f{1}{2}(\alp-\bt)|\ge\epsl$.
We thus have
\Al{\label{ceguorbgwogb}
\sqrt{ \brak{ U^{\ot n}\phi^+U^{\dagger,\ot n},\phi^+_V } } &= \Big|\f{1}{n-1}\big\langle U_{\f{n}{2}-1},V_{\f{n}{2}-1} \big\rangle\Big| \notag \\
&=\Big|\f{1}{n-1} \f{\ee^{\ii(n\alp+\bt)}-\ee^{\ii(\alp+n\bt)}}{\ee^{\ii\alp}-\ee^{\ii\bt}} \Big|\notag \\
&=\Big|\f{1}{n-1} \f{\sin\f{n-1}{2}(\alp-\bt)}{\sin\f{1}{2}(\alp-\bt)} \Big| \notag \\
&\le \Big|\f{1}{n-1} \f{1}{\sin\f{1}{2}(\alp-\bt)} \Big| \notag\\
&\le \f{1}{(n-1)\epsl} \,,
}
where the second equality used the Weyl character formula~\eqref{owdhbeobhro} for irrep $\Hc_{\f{n}{2}-1}$.
We now take $n=\f{\sqrt3}{\epsl}+1=O(\epsl^{-1})$ so that the soundness condition $\brak{ U^{\ot n}\phi^+U^{\dagger,\ot n},\phi^+_V }\le1/3$ for $\x{dist}(U,V)\ge\epsl$ is satisfied.
\end{proof}


When both single-qubit unitaries are unknown, we have the following.
\Thm{\label{thm:vnewrhod2}
	Given access to two unknown single-qubit unitaries $U$ and $V$, there exists a tester with perfect completeness that distinguishes whether $U$ equals $V$ up to a phase or $\x{dist}(U,V)\ge\epsl$ with $O(\epsl^{-1})$ uses of $U$ and $V$, without using any ancilla system.
}
\begin{proof}
Similar to the proof of Theorem~\ref{thm:vnewrhod}, we first decompose the space as $(\Cb^2)^{\ot n}=\Oplus_j \Hc_j\ot\Kc_j$ where $\Hc_j$ and $\Kc_j$ are irreps of $\Ux(2)$ and $S_n$ respectively.
Denote by $\ket{\phi^+}$ the maximally entangled state in $\Hc_{\f{n}{2}-1}\ot\Kc_{\f{n}{2}-1}$.
Then we apply the swap test to $U^{\ot n}\ket{\phi^+}$ and $V^{\ot n}\ket{\phi^+}$.
Repeat the above procedure, and report that the two unitaries are equal up to a phase if and only if the swap test accepts twice.
When $U$ and $V$ are equal, the tester reports correctly with certain.
When $\x{dist}(U,V)\ge\epsl$, since $\big|\brak{ \phi^+|U^{\dagger,\ot n}V^{\ot n}|\phi^+ }\big|\le \f{1}{(n-1)\epsl}$ by~\eqref{ceguorbgwogb}, the tester reports incorrectly with probability less than $\big(\f{1}{2}(1+\f{1}{(n-1)^2\epsl^2})\big)^2$, which is in turn less than $1/3$ if we take $n=\f{3}{\epsl}+1$.
\end{proof}

In order to deal with the higher dimensional case, an upper bound on $|\tr U_\ld|/d_\ld$ is needed given $|\tr U|/d\le 1-\epsl^2$, where $U_\ld$ is representation matrix of $U$ on irrep $\Hc_\ld$ for appropriate $\ld\in\x{Par}(n)$ and $d_\ld$ is the dimension of $\Hc_\ld$.

\Lem{\label{lemma-blowup}
	Let $\ld=(d-1,d-2,\dots,0) n/\binom{d}{2}\in\x{Par}(n,d)$ and $s=n/\binom{d}{2}+1$.
	If $\f{1}{d}|\brak{U,V}|\le1-\epsl^2$ for $U,V\in\Ux(d)$, then
	\Eqn{
	\f{1}{\dim\Hc_\ld}|\brak{U_\ld,V_\ld}| \le \Big(\f{2}{s\epsl}\Big)^m
	}
	for some positive $m$.
}
\begin{proof}
The dimension of the irrep $\Hc_\ld$ is
\Eq{\label{dimvowehgwbf}
	\dim\Hc_\ld = s^{d(d-1)/2} \,.
}
The character for $\Hc_\ld$ is
\Eqn{
	\chi_\ld^L(\x{diag}(x_1,\dots,x_d)) = \f{ \prod_{1\le j<k \le d}(x_k^s-x_j^s) }{ \prod_{1\le j<k \le d}(x_k-x_j) } \,.
}

Since $\s{R}_\ld:U\mapsto U_\ld$ is a unitary representation of $\Ux(d)$, we have $U_\ld V_\ld=(UV)_\ld$ and $(U_\ld)\dg=(U\dg)_\ld$.
Thus $\brak{U_\ld,V_\ld}=\brak{(U\dg V)_\ld,\1}$.
It suffices to consider the case $V=\1$.

Consider a unitary $U$ having eigenvalues $\ee^{\ii\tht_k}$ with $0\le \tht_k<2\pi$ for each $k\in[d]$.
We have
\Al{\label{vjengornfkdne}
	|\tr U|^2 = \bigg|\sum_{k=1}^d \ee^{\ii\tht_k}\bigg|^2 &=  d+2\sum\nolimits_{1\le j<k\le d}\cos(\tht_k-\tht_j) \nonumber\\
	&= d^2-4\sum\nolimits_{1\le j<k\le d}\sin^2\f{1}{2}(\tht_k-\tht_j)  \,,
}
while
\Al{\label{vjenfkdnewiojrnwrg}
| \tr U_\ld | &= \prod_{1\le j<k \le d} \bigg| \f{ \ee^{\ii s\tht_k}-\ee^{\ii s\tht_j} }{ \ee^{\ii \tht_k}-\ee^{\ii \tht_j} } \bigg| \nonumber \\ 
&= \prod_{1\le j<k \le d}\bigg| \f{ \sin\f{s}{2}(\tht_k-\tht_j) }{ \sin\f{1}{2}(\tht_k-\tht_j) } \bigg| \,.
}

Before proceeding with the proof we now show that
\Eq{\label{wohvoeaufcvd}
\Big|\f{\sin sx}{\sin x}\Big|\le s
}
holds for any odd positive integer $s$ and any $|x|\le \pi $.
Indeed, it suffices to consider the case $0\le \sin x\le\f{1}{s}$, and~\eqref{wohvoeaufcvd} follows by noticing that the function $\f{\sin sx}{s\sin x}$ is even and is symmetric about the line $x=\f{\pi}{2}$.
When $0\le \sin x\le\f{1}{s}$, since $\sin x\ge \f{2}{\pi}x$, we have $sx\le\f{\pi}{2}$.
It follows that for any $s\ge1$, we have $\cos x\ge\cos sx$, and thus $s\sin x\ge\sin sx$, completing the proof of~\eqref{wohvoeaufcvd}.

As for the soundness condition, when $\f{1}{d}|\tr U|\le 1-\epsl^2$, by~\eqref{vjengornfkdne} we have
\Eqn{
	\sum_{1\le j<k\le d}\sin^2\f{1}{2}(\tht_k-\tht_j)\ge \f{1}{4}d^2(2\epsl^2-\epsl^4) \,.
}

It follows that there exists $(j,k)$ satisfying
\Eqn{
\sin^2\f{1}{2}(\tht_k-\tht_j)\ge \f{d(2\epsl^2-\epsl^4)}{2(d-1)} \,,
}
and let $m$ be the number of such pairs in totally $\binom{d}{2}$ pairs.
For any such pair,
\Eq{\label{vekrghoer}
\f{ |\sin\f{s}{2}(\tht_k-\tht_j)| }{ |\sin\f{1}{2}(\tht_k-\tht_j)| } \le \f{ 1 }{ |\sin\f{1}{2}(\tht_k-\tht_j)| } \le \sqrt{ \f{2(d-1)}{d(2\epsl^2-\epsl^4)} }\le \f{2}{\epsl}
}
for $\epsl\le1$.
On the other hand, there are $\binom{d}{2}-m$ pairs of $(j,k)$ satisfying $\sin^2\f{1}{2}(\tht_k-\tht_j) < \f{d(2\epsl^2-\epsl^4)}{2(d-1)}$.
For any such pair, since $|\f{1}{2}(\tht_k-\tht_j)| \le\pi$, by~\eqref{wohvoeaufcvd} we have
\Eq{\label{vekrghoer222}
\f{ |\sin\f{s}{2}(\tht_k-\tht_j)| }{ |\sin\f{1}{2}(\tht_k-\tht_j)| }\le s \,.
}

Therefore, combining~\eqref{dimvowehgwbf},~\eqref{vjenfkdnewiojrnwrg},~\eqref{vekrghoer} and~\eqref{vekrghoer222},
\[\f{1}{\dim\Hc_\ld}|\tr U_\ld|\le s^{-\binom{d}{2}} (2/\epsl)^m s^{-m+\binom{d}{2}} =\Big(\f{2}{s\epsl}\Big)^m \,.
\]
\end{proof}


\Thm{\label{dgwegnkwebngjbw}
	Given access to an unknown unitary $U\in\Ux(d)$ and a known or unknown unitary $V\in\Ux(d)$, there exists a tester with perfect completeness that distinguishes whether $\x{dist}(U,V)=0$ or $\x{dist}(U,V)\ge\epsl$ with $O(d^2/\epsl)$ uses of unitaries.
}


\begin{proof}
Consider the partition $\ld=(d-1,d-2,\dots,0) n/\binom{d}{2}$, and denote $s:=n/\binom{d}{2}+1$.
The proof follows similar approach used in Theorems~\ref{thm:vnewrhod} and~\ref{thm:vnewrhod2}.

By noticing that $\ld/n$ is independent of $n$, it follows from~\eqref{vmognreogbjgt2} and~\eqref{vmognreogbjgt} that when $\epsl$ is so small that $n$ is much larger than $d$, the dimension of $\Kc_\ld$ is exponential in $n$ while the dimension of $\Hc_\ld$ is polynomial in $n$, thus $\dim\Kc_\ld$ is larger than $\dim\Hc_\ld$.

Denote by $\ket{\phi^+}$ the maximally entangled state in $\Hc_\ld\ot\Kc_\ld$, and denote $\ket{\phi_U^\x{m}}:=U^{\ot n}\ket{\phi^+}$ and $\ket{\phi_V^\x{m}}=V^{\ot n}\ket{\phi^+}$.
When $U$ is unknown and $V$ is given, perform a measurement $\{\phi_V^\x{m},\1-\phi_V^\x{m}\}$ on $\phi_U^\x{m}$, and repeat, and accept if and only if the first outcome occurs.
When $U$ and $V$ are both unknown, perform a swap test for $\phi_U^\x{m}$ and $\phi_V^\x{m}$, and repeat.

The completeness condition is obvious.
When $\x{dist}(U,V)\ge\epsl$, using Lemma~\ref{lemma-blowup}, the soundness condition is satisfied by taking $s$ to be the smallest odd integer larger than $6/\epsl$, for which $n\le \binom{d}{2}(\f{6}{\epsl}+1)=O(d^2/\epsl)$.
\end{proof}


For the asymptotic case where the value of $\epsl$ is small, $n$ will be so large that the irrep $\Hc_\ld$ has smaller dimension than its multiplicity space $\Kc_\ld$ does.
Thus there exists a maximally entangled state $\phi^+$ on $\Hc_\ld\ot\Hc_\ld$ which is a subspace of $\Hc_\ld\ot\Kc_\ld$.
If $\epsl$ is not small enough and thus $n$ is not large enough, one can introduce a reference system of dimension $\f{\dim\Hc_\ld}{\dim\Kc_\ld}$ to make them have equal dimension \cite{chiribella2005optimal}.
The dimension of the ancilla system, however, has been exponentially decreased compared with the method using Choi states directly.
Since it is common to deal with low-dimensional quantum systems in quantum computing under current technology, our method exhibits advantage in practice.

For the certification of identity of qudit unitaries, we have considered the irrep corresponding partition $\ld=(d-1,d-2,\dots,0)n/\binom{d}{2}$.
One issue that would be worth investigating is whether the irreps of unitary group used in this chapter are optimal compared with other irreps.

Using Theorem~\ref{dgwegnkwebngjbw}, one can also efficiently test whether one unitary is identity, and whether two unitaries are inverse to each other.
It is known that many properties of quantum channel can be reduced to testing properties of quantum state via the Choi-Jamio{\l}kowski isomorphism, but by employing this approach one usually needs to use quantum channels too many times and also to introduce extra ancilla system.
In this section we used group representation to test properties of unitaries more sample-efficiently without using ancilla system.
Our approach may be promising in the property testing of noisy quantum operation and measurement to achieve better performance.
It remains to give a lower bound on the query complexity for the unitary certification.



\section*{Acknowledgements}
This work was supported by 
National Natural Science Foundation of China (Grant No.\ 62102388),
Innovation Program for Quantum Science and Technology (Grant No.\ 2021ZD0302902),
Fundamental Research Funds for the Central Universities (Grant No.\ WK2150110023),
and Anhui Initiative in Quantum Information Technologies (Grant No.\ AHY150100).

\appendix
\section{Upper bound on query complexity}

Following the work of \cite{kahn2007fast}, this appendix gives an upper bound on the query complexty of ancilla-free quantum unitary tomography: There is an algorithm that requires $O(d^2\epsl^{-1/2})$ uses to estimate a unitary process acting on a $d$-level quantum system to error $\epsl$ in infidelity.

For estimating an unknown unitary in $\SU(d)$, the asymptotic dependence of error $R$  on number $n$ of particles used was derived in \cite{kahn2007fast}, that is, $R=\Tht(n^{-2})$ for constant $d$.
We now upper bound the error $R$ for $n$ and $d$ both sufficiently large based on the strategy used in \cite{kahn2007fast}.
For the sake of completeness we first introduce the strategy and derive $R$ in terms of $n$ and $d$.

Let $\Nb$ denote the set of all nonnegative integers.
For $n,d\in\Nb$, let $\Pc_d(n)$ denote the set of partitions of $n$ with length at most $d$, i.e., $\Pc_d(n):=\{\ld=(\ld_1,\dots,\ld_d)\in\Nb^d:\ld_1\ge\cdots\ge\ld_d,\sum_{i=1}^d\ld_i=n\}$.
A partition $\mu=(\mu_1,\dots,\mu_d)$ is called {\em strict} if $\mu_1>\cdots>\mu_d>0$; denote $\Pc_d^\x{s}(n):=\{\ld\in\Pc_d(n):\ld\x{ is strict}\}$.


By Schur-Weyl duality, $(\Cb^d)^{\ot n}=\Oplus_{\ld\in\Pc_d(n)}\Hc_\ld\ot\Kc_\ld$ where $\Hc_\ld$ and $\Kc_\ld$ are irreps of unitary group $\SU(d)$ and symmetric group $S_n$ respectively corresponding to partition $\ld$.
Let $d_\ld$ and $\dim\ld$ denote the dimensions of $\Hc_\ld$ and $\Kc_\ld$ respectively.
Let $\ket{\phi_\ld^+}=\f{1}{\sqrt{d_\ld}}\sum_{i=1}^{d_\ld}\ket{i,i}$ be a maximally entangled state of rank $d_\ld$.
For any unknown unitary $U^{\ot n}$, let the input state be $\ket{\psi}=\Oplus_{\ld\in\Pc_d(n)}c_\ld\ket{\phi_\ld^+}$ with $\sum_\ld |c_\ld|^2=1$.
When $d_\ld>\dim \ld$ for some $\ld$, we set $c_\ld=0$.

Let $M(U)=U^{\ot n}(\Oplus_\ld d_\ld^2\phi_\ld^+)U^{\dagger,\ot n}
= \Oplus_\ld d_\ld^2(U_\ld\ot\1)\phi_\ld^+(U_\ld\dg\ot\1)$.
Then $\Pi:=\int M(U)\dd U=\Oplus_\ld \1_{\Hc_\ld}\ot \1_{\Hc_\ld}$ is a projector.
The POVM elements $M(U)\dd U$, together with $\1-\Pi$, constitute a complete POVM.

The point risk, i.e., the supremum expected error $R(n,d):=\sup_U \E \Dt(U,\widehat U)$, of the above strategy is with respect to the metric
\Eq{
\Dt(U,\widehat U):=1-|\f{\tr(U\dg \widehat U)}{d}|^2,
}
which is called infidelity in this appendix.
Now $\Pr(\widehat U|U)=\brak{U^{\ot n}\psi U^{\dagger,\ot n},M(\widehat U)}
=\Pr(U\dg \widehat U|\1)$.
Thus $\E\Dt(U,\widehat U)=\int\Pr(\widehat U|U)\Dt(U,\widehat U)\dd\widehat U=\E\Dt(\1,\widehat U)$, which is independent of $U$.
\Aln{
	R(n,d) &= \int \Pr(\widehat U|\1)\Dt(\1,\widehat U) \dd\widehat U\\
	&\le 1-\int |\brak{\psi|\oplus_\ld d_\ld(U_\ld\ot\1)|\phi_\ld^+}|^2|\tr U/d|^2 \dd U\\
	&= 1-\f{1}{d^2}\int \Big|\sum\nolimits_\ld c_\ld^* d_\ld\brak{\phi_\ld^+|(U_\ld\ot\1)|\phi_\ld^+}\Big|^2 |\tr U|^2 \dd U \\
	&= 1-\f{1}{d^2}\int \Big|\sum\nolimits_\ld c_\ld^* \chi_\ld(U) \Big|^2  |\tr U|^2 \dd U \\
	&=1-\f{1}{d^2}\int \bigg|\sum_{\mu\in\Pc_d(n+1)}\sum_{\ld\nearrow\mu}c_\ld^*\chi_\mu(U) \bigg|^2 \dd U \\
	&=1-\f{1}{d^2} \sum_{\mu\in\Pc_d(n+1)}\bigg|\sum_{\ld\nearrow\mu}c_\ld \bigg|^2 \,,\\
	&\le1-\f{1}{d^2} \sum_{\mu\in\Pc_d^\x{s}(n+1)}\bigg|\sum_{\ld\nearrow\mu}c_\ld \bigg|^2 \,,
}
where in the first line the variable $\widehat U$ of integration changes to $U$ for brevity, and the last 2nd line uses orthogonality relation for characters of irreps.

For partition $\ld\in\Pc_d(n)$, denote $p_\ld:=p_1\cdots p_d$  where $p_i:=\ld_i-\ld_{i+1}$ (note that $\ld_{d+1}=0$); for partition $\mu\in\Pc_d(n+1)$, denote $q_\mu$ in the same way.
Let $c_\ld=p_\ld/a$ where $a=\sqrt{\sum_{\ld\in\Pc_d(n)} p_1^2\cdots p_d^2}$ is a normalization constant.
It can be shown that $c_\ld=0$ whenever $d_\ld>\dim \ld$.
Thus
\Al{\label{eq-1}
	R(n,d) &\le 1-\f{ \sum_{\mu \x{ strict}}\big(\sum_{\ld\nearrow\mu}p_\ld \big)^2 }{  d^2a^2 } \notag\\
	&= 1-\f{ \sum_{\mu \x{ strict}}\big(\sum_{i=1}^d p_{\mu-e_i} \big)^2 }{  d^2a^2 } \notag\\
	&= 1-\f{ \sum_{\mu \x{ strict}} (dq_\mu+\sum_{i=1}^d r_i)^2 }{ d\sum_{\mu\in\Pc_d(n+1)}\sum_\ld p_\ld^2 } \notag\\
	&\le 1-\f{ \sum_{\mu \x{ strict}} (dq_\mu+\sum_{i=1}^d r_i)^2 }{ d\sum_\mu\sum_{i=1}^d p_{\mu-e_i}^2 } \notag\\
	&= 1-\f{ \sum_{\mu \x{ strict}} (dq_\mu+\sum_{i=1}^d r_i)^2 }{ d\sum_\mu\sum_{i=1}^d (q_\mu+r_i)^2 } \,.
}
Here $r_i:=q_{\backslash i-1}-q_{\backslash i}-q_{\backslash i,i-1}$, where $q_{\backslash i}:=q_\mu/q_i$ and $q_{\backslash i,i-1}:=q_\mu/(q_iq_{i-1})$ expect that $q_{\backslash0}=q_{\backslash 1,0}=0$.
Omitting the higher order we take $r_i=q_{\backslash i-1}-q_{\backslash i}$, then $\sum r_i=-q_{\backslash d}$.
For any partition $\mu\in\Pc_d(n+1)$ with $q_i:=\mu_i-\mu_{i+1}$, denote $x_i:=q_i/(n+1)$ and similarly for $x_\mu$ and $x_{\backslash i}$.
Denote $T_k:=1-\sum_{j=1}^k jx_j$ for $k\in[d]$; notice that $T_d=0$.
It follows from \eqref{eq-1} that
\Eq{\label{eq-6}
	R(n,d) \le 1-\f{ \sum_{\mu \x{ strict}} ((n+1)dx_\mu-x_{\backslash d})^2 }{ d\sum_\mu\sum_{i=1}^d ((n+1)x_\mu+x_{\backslash i-1}-x_{\backslash i})^2 } \,.
}

As $n$ goes to infinity, the Riemann sums in the above inequality converge to Riemann integrals and do not depend on whether $\mu$ is strict.
Then the r.h.s.\ of \eqref{eq-6} converges to
\Eq{\label{eq-8}
	1-\f{ \int ((n+1)dx_\mu-x_{\backslash d})^2\dd x }{ d\int\sum_{i=1}^d ((n+1)x_\mu+x_{\backslash i-1}-x_{\backslash i})^2\dd x } 
	=
	\f{ d\int\sum_{i=1}^d(x_{\backslash i-1}-x_{\backslash i})^2\dd x-\int x_{\backslash d}^2\dd x }{ d\int\sum_{i=1}^d ((n+1)x_\mu+x_{\backslash i-1}-x_{\backslash i})^2\dd x } \,.
}
where all integrals are over the domain $\{(x_1,\dots,x_{d-1}):x_i\ge0\x{ for each }i\x{ and }T_{d-1}\ge0\}$ and $\x{d}x\equiv\x{d} x_1\cdots \x{d} x_{d-1}$.




The dependence of risk on $n$ can be seen from \eqref{eq-8}: $R(n,d)=O(n^{-2})$ for constant $d$ \cite{kahn2007fast}.
All integrals in \eqref{eq-8} can be calculated via recursion in the following way:
\Aln{
	\int x_\mu\dd x &= \f{1}{d^2}\int x_1^2\cdots x_{d-1}^2 T_{d-1}^2\dd x_1\cdots \dd x_{d-1} \\
	&=\f{1}{d^2} \f{B(3,3)}{(d-1)^3} \int x_1^2\cdots x_{d-2}^2 T_{d-2}^5\dd x_1\cdots \dd x_{d-2} \\
	&=\f{1}{d^2} \f{B(3,3)B(3,6)}{(d-1)^3(d-2)^3} \int x_1^2\cdots x_{d-3}^2 T_{d-3}^8\dd x_1\cdots \dd x_{d-3}
}
where $B(\ap,\bt):=\int_0^1 t^{\ap-1}(1-t)^{\bt-1}\dd t$ is the beta function, and finally,
\Als{\label{eq-9}
	\int x_\mu\dd x &= \f{B(3,3)B(3,6)\cdots B(3,3(d-1))}{d^2((d-1)!)^3} \\
	&=\f{2^d}{d^2((d-1)!)^3(3d-1)!} \,.
}
Similarly we have
\Al{
	\int x_{\backslash i}^2\dd x &= \f{2^{d-1}i^2}{d^2((d-1)!)^3(3d-3)!}, \forall i\in[d]\,,\\
	\int x_{\backslash i}x_{\backslash i+1}\dd x &= \f{2^{d-2}i(i+1)}{d^2((d-1)!)^3(3d-3)!}, \forall i\in[d-1]\,,\\
	\int x_\mu^2/x_d\dd x &= \f{2^{d-1}}{d((d-1)!)^3(3d-2)!} \label{eq-10}\,.
}

Invoking \eqref{eq-9}-\eqref{eq-10}, the numerator and denominator of \eqref{eq-8} are equal to $\f{1}{3}\f{2^{d-1}}{((d-1)!)^3(3d-1)!} (d^2-1)(3d-2)(3d-1)$ and $\f{2^{d-1}}{((d-1)!)^3(3d-1)!}(2(n+1)^2-2(n+1)(3d-1)+\f{1}{3}(3d-2)(3d-1)(d^2+2))$, respectively. Thus \eqref{eq-8} is equal to
\Eqn{
	\f{ (d^2-1)(3d-2)(3d-1) }{ 6(n+1)^2-6(n+1)(3d-1)+(3d-2)(3d-1)(d^2+2) } \,,
}
which is less than $\epsl$ for large $d$ if $n=Cd^2\epsl^{-1/2}$ for some constant $C$.

\bibliographystyle{apsrev}
\bibliography{../bib}
\end{document}